# Unique Room Temperature Light Emitting Diode Based on 2D Hybrid Organic-Inorganic Low Dimensional Perovskite Semiconductor

Anastasia Vassilakopoulou, Dionysios Papadatos and Ioannis Koutselas*

Materials Science Department
School of Natural Sciences
University of Patras, Greece, 26504
*ikouts@upatras.gr

**Keywords**: Hybrid organic-inorganic low dimensional semiconductor; exciton; light emitting diode; Photoluminescence; Electroluminescence

**Abstract**

A Light Emitting Diode (LED) based on a two dimensional (2D) Hybrid Organic-Inorganic Semiconductor (HOIS) functioning at room temperature is demonstrated. LED is fabricated by depositing a layer of (4-fluorophenethylamine-H)$_2$PbI$_4$ on an ITO substrate. It is the first time that such a room temperature excitonic LED based on 2D lead iodide perovskite is presented, as well as for its simplicity, low cost, ambient conditions fabrication and functioning at 5-10V. The newly introduced class of 3D HOIS perovskite LEDs, is now broadened with the implementation of the 2D HOIS. Novel functionalities can be realized since it is now possible to access the hybrid's 2D semiconductor advantageous properties, such as increased excitonic binding energy, oscillator strength and peak wavelength tunability.

## 1. Introduction

Low Dimensional (LD) Hybrid Organic-Inorganic Semiconductors (HOIS) exhibit novel optical quantum phenomena, due to the enhanced quantum and dielectric confinement of their LD excitons [1,2]. HOIS are capable of spanning dimensionalities, as far as their semiconducting inorganic network is concerned, from 3D to 0D, including intermediate dimensionalities such as the so called quasi-1D and quasi-2D HOIS [2]. This natural, self-assembling and low cost class of LD semiconductors exhibits useful optoelectronic properties [3-8] comparable to those of the artificial semiconductor class, for which a number of issues may be addressed for their optimization, as in [9-11]. Light Emitting Diodes (LEDs) based on 2D [12-14] or 3D [15-18] HOIS are such device examples. LD HOIS exhibit excitonic states of increased binding energy ($E_b$) and oscillator strength ($f_{exc}$), while the associated excitonic Optical Absorption (OA) and accompanying Photoluminescence (PL) peak can be tuned by varying their stoichiometry or structure [2]. The increased $E_b$ is mainly due to the dielectric mismatch of the alternating inorganic and organic subparts of the HOIS [1,3]. The inorganic part is usually the semiconductor, composed of a metal halide unit network, while the organic part is formed from an appropriate self-assembled packing of electrooptically inactive amines. In the 2D HOIS, which self-assemble as a superlattice, the thickness of the repeating active layer is c.a. 0.6nm, while that of the organic layer is about 1-3nm, depending on the choice of the organic molecule. In the past, a wide set of combinations of inorganic and organic parameters have led to the synthesis of HOIS with a variety of properties and applications. Such examples are HOIS with controllable excitonic peaks covering the ultraviolet-visible region [2], thin film transistor gate materials comparable to amorphous Si [19-21] or HOIS exhibiting extremely interesting energy transfer optical phenomena [22,23]. Finally, the 3D and quasi-2D lead halide HOIS have been successfully employed for solar cell applications [24,25].

In this work we report a new simple method for fabricating a 2D HOIS based LED, which operates at room temperature (RT). Initial results reported here show the advantages of the hybrid organic-inorganic lead halide perovskites in performing basic and applied research, where stable 2D excitons are required at room temperature. Until now no work has shown that a 2D HOIS based LED can function at room temperature, to our knowledge with the exception of the oleylamine based HOIS reported in *ref.* [14], where the 4-

fluorophenethylamine case was also mentioned. In *ref.* [14], the oleylamine based prototypes in some cases required high voltages, i.e. above 10V, in order to observe Electroluminescence (EL) depending on the film thickness and the halogen content of the semiconductor. Also, the main EL peak had significant contribution from strong Stokes shifted emission *wrt* to the excitonic OA peak. The latter could be attributed to impurities in the organic amine leading to nanostructures with red shifted or defect-like excitonic states. In HOIS, with multiple excitonic states, including defect-like states, energy is being transferred from higher energy excitonic states to those of lower energy [22,23]. Lately, this has also been observed in LEDs based on such HOIS and is reported elsewhere.

Also, in a recent publication [26], following this work's submission, EL of a 2D HOIS based on phenethylamine, lead bromide and a different fabrication method than the one presented here, has been reported, showing the immense capabilities of the perovskite 2D systems. It is conjectured that future studies of the perovskite systems with more detailed micro and/or nano photoluminescence experiments, taking into account physical parameters, may yield nanoscale LEDs of enhanced performance [27,28].

## 2. Experimental

**2.1 Chemicals:** 4-Fluorophenethylamine (abbr. FpA, 98%), Acetonitrile (abbr. AcN, CHROMASOLV® Plus, for HPLC, ≥99.9%), N,N'-Dimethylformamide (abbr. DMF, 99.8%), Hydriodic acid (ACS reagent, ≥47.0%), Lead (II) iodide (99.999% trace metals basis), Gallium–Indium eutectic (≥99.99% trace metals basis), Indium Tin Oxide (ITO) coated glass (square, surface resistivity 15-25 Ω/sq) were obtained from Sigma Aldrich and used as such with no further purification.

**2.2 Synthesis:** Synthetic route was followed as reported in [29,30] and analogous to those in [4,21]. In summary, 8mmole of FpA were mixed, at 48$^o$C, in 5 ml of AcN and treated slowly with 8mmole of HI (FpA-H denotes the protonated molecule); 4mmole PbI$_2$ was added in 5ml of AcN and treated with 9.5mmole HI at 48°C. The second solution was slowly added, under mixing, to the first solution. Slow cooling gave rise to red-orange crystals, which have been dried extremely well from solvent. These semiconductor crystals show strong PL even under ambient room light, with a characteristic green coloring. LEDs were constructed by a simple doctor-blade method where 20μL of solution (60mg of $(FpA-H)_2PbI_4$ in 200μl DMF) was deposited on the ITO surface while it was heated at 50°C for a short period. Solutions of higher concentration have also led to functional LEDs. As preliminary result, it has been observed that synthesis from higher protonated amine content solutions also yield functional LEDs. For example, instead of using the above mentioned precursor molar ratio amine:acid:$PbI_2$ of 2:4.3:1, sample DP268 will refer to material synthesized from solution with ratio of 8:6.3:1, which allows better function of the LEDs. This specific sample was not rinsed for removal of unreacted components.

The ITO coated substrates were cleaned by immersion in piranha solution for 4 minutes and rinsed with 18MΩ water. No electron and hole injection layers were employed above or below the semiconductor layer, although this is possible. LED's top contact was made using the Ga/In alloy tipped aluminum electrode. The DMF solution used to prepare LEDs was also stored in ambient air for one day to test for possible material oxidation, which was not observed in macroscopic optical measurements but is speculated that it does occur.

**2.3 Characterization:** Structural and spectroscopic information for the 2D HOIS is obtained with instrumentation as reported before [29]. Recorded video and images of the LED turned on, were obtained by a standard computer camera, yet easily visible by naked eye. Captured data were recorded for forward bias only, of about 5-10V, as reverse bias did not provide light emission. EL spectra were recorded by modifying a θ-Metrisis optical profilometer equipped with an Ocean Optics polychromator CCD, with non-cooled detector. I-V characteristic curves were recorded using a Keithley 6517A measuring current at specified internally generated voltages, where a sub-millimeter radius of Cu wire, dipped in

Ga/In and in the form of a spring, slightly touched the surface of the active semiconducting material. In some cases, the value of currents recorded for the I-V (supplementary section, Fig. S5) is smaller compared to those recorded when the device is turned on using a Ga/In dipped banana connector, probably due to the larger contact area of the Ga/In liquid banana tip and possibly increased leakage current values.

## 3. Results and Discussion

A 2D HOIS based on lead iodide and 4-fluorophenethylamine has been synthesized, characterized and implemented as a LED device. Under appropriate reaction conditions, crystalline 2D HOIS with formula $(FpA-H)_2PbI_4$ were produced. Mitzi et al. had used initially FpA for synthesizing low melting point temperature HOIS [4]. The final material here is comprised of dark orange crystals, yielding strong photoluminescence when excited with UV, e.g. 404nm laser, which is due to the first exciton recombination occurring at the Γ-point, visible by naked eye even at RT [31].

Fig. 1 shows typical powder XRD patterns of $(FpA-H)_2PbI_4$ and that of a similar synthesis, where a threefold increase of the needed protonated amine was used (DP268) along with some unreacted amine. Both samples exhibit peaks at the same 2θ positions, however, the DP268 sample has relatively less intense peaks at low angles as the extra amine probably excludes the formation of large number of 2D platelets. The peaks for both types of samples are consistent with a monoclinic unit cell, where the long axis is the sum of two FpA molecules and two Pb-I bonds, c.a. 16.5Å. The low angle peaks are due to the superlattice formed by the alternating organic and inorganic layers [29].

Optical properties of the 2D $(FpA-H)_2PbI_4$ were studied by means of OA and PL spectroscopy, as presented in Fig. 2. More specifically, the inset in Fig. 2 shows an ITO glass coated with thick deposits of $(FpA-H)_2PbI_4$ from DMF solution, which was used as such for EL measurements. It is centrally irradiated with a 404nm laser, thus, observing the strong green exciton luminescence [2]. Fig. 2a, shows the OA spectra for this particular sample, while Fig. 2b shows the corresponding PL spectra under 300nm excitation. Due to UV absorption of the ITO glass as well as sample's non-ideal thickness, a more detailed OA spectra of thin deposits on quartz substrate is provided in Fig. 2c. The excitonic OA peak occurs at 513nm, with an onset of band gap absorption appearing at 470nm, indicating an excitonic binding energy of at least 220meV. The spectra in Fig. 2a differs from that in Fig. 2c in the fact that the first shows broad absorption peaked at 527nm, while its absorption curve continues up to 540nm. Moreover, the OA spectra in Fig. 2c exhibits a characteristic peak at 380nm, due to the $PbI_6$ octahedron energy levels perturbed by the neighboring octahedra [2]. If scattering effects should be neglected due to film's thickness, some of the differences between the two OA spectra can be understood as HOIS often accommodate slightly varying structures depending on the solvent, the crystallization process and the time afforded for the solvent's drying. Such examples are being provided in the supplementary section as Fig. S1 and S2, regarding the $(FpA-H)_2PbI_4$ OA and PL spectra, respectively, of the compound used in this work as synthesized and as re-crystallized from AcN or DMF. Fig. 2b exhibits PL peak at 526nm which is Stokes shifted with a maximum of 10nm (64meV), indicating the good crystalline state of the material as deposited on the ITO substrate. In addition, Fig. 2b spectra exhibits an almost undetectable shoulder at about 540nm. In the supplementary section (Fig. S2), PL spectra of the same compound as synthesized and as re-crystallized from various solvents show the existence of two photoluminescence mechanisms, with peaks at 527nm and 545nm. The peak at 527nm is associated with free exciton recombination while the peak at 545nm may be linked with bound excitons, defects or some material phase change. The latter peak coincides with a small shoulder in some of the corresponding OA spectra in Fig. S1, at c.a. 540nm. Should such a phase change exist, it certainly does not appear in the XRD patterns, therefore, if existing it must be only composed of nanoscale perovskite particles. Second low energy PL peak disappears when the material is being re-crystallized from AcN or DMF while the intensity of the associated shoulder in the OA spectra is decreased as well.

The film, whose optical properties are shown in Fig. 2a and 2b, shows strong, naked-eye visible, EL when electrons are injected to it via a Ga/In alloy electrode and holes are injected via the ITO electrode. Fig. 3 shows images obtained from videos of this type of LEDs functioning at 8V, while voltages between 5-10V usually suffice for operation. Video of LED's operation as well as wavelength analyzed EL emission can be found as supplementary data (Fig. S3). The EL exhibits a peak at 555nm and is rather broad.

The EL peak is red shifted with respect to the OA and PL peaks of Fig. 2 while its spectra is asymmetrical with respect to its peak. XRD, OA and PL spectra present only indiscernible differences between films of (FpA-H)$_2$PbI$_4$ as re-crystallized from DMF and films of the starting material as crystallized from AcN. Exception to that constitutes the diminishing of the initially observed low energy PL peak after the DMF recrystallization. Therefore, it was hypothesized that an interesting recombination process leads to the observed EL shift, which is not related to the bulk of the material. Similar red shift has been observed and explained, for example in *refs*. [32,33], attributing it to various localized phenomena or high electric field polarization to which the interface band bending should be added.

In order to test whether bound excitons, defects or some sort of energy transfer [22,23] is the origin of the red shift, PL spectra of two semiconductor test films, a thin and a thick, have been acquired in two different configurations (front face excitation and emission, front face excitation with back face emission as seen in schematic within Fig. S4). PL data for these two mentioned configurations are presented in the supplementary section (Fig. S4), where it can be observed that the two test films show shifted PL peaks when the PL emission is measured from the front or the back side, while the excitation is on the front side of the film. Thin film shows smaller PL shift than the thick film. Specifically, the back emitted PL of the thick film exhibits a red shift of c.a. 20nm *wrt* to PL peak emitted from the front face. This red shift along with the fact that the back emitted light distribution is asymmetrical (Fig. S4) while it lacks front face's PL peak, is attributed to self-absorption. The fact that there is a detected peak in the back face PL light, instead of a monotonic decreasing signal, may be attributed to re-emission of the self-absorbed radiation, probably coupled to various low energy emission mechanisms inherent in the film.

Finally, SEM images of the as prepared (FpA-H)$_2$PbI$_4$ and of that deposited, as described in the experimental section, on substrates are shown in Fig. 4. It can be observed that initially the HOIS powder is composed of thin square like platelets, some of the order of 100μm x 100μm, while that of the deposited material, which shows EL, is composed of a random packing of much smaller 2D platelets. In all cases shown in Fig. 4, EDX measurements showed that the molar ratio of I:Pb was 4, as expected. Initially, as in *ref.* [14], it was thought that in order to operate a LED, current would have to find a path through the non-conducting amine layers, which can be a serious problem as the 2D HOIS tend to preferentially align the organic 2D layers in parallel to the ITO substrate. Therefore, a random packing of small 2D platelets may allow, especially close to the ITO surface, electrons and holes to move between the anode and cathode as these hop from one nanocrystal to another. At this time it is not known if an assortment of such closely packed nanocrystals can show phenomena as in [23] where super-radiance is observed.

It is evident that the simple methodology presented here allows the fabrication of a LED device operating at RT, which also readily functions when cooled at 77K. Until this report's writing, LEDs formed with the same method but based on 2D HOIS with different short amines, have not exhibited emission of light. It is conjectured, according to the structure solution and analysis performed by Mitzi et al. for the SnI$_4$ analogue [4], that the large distances of the fluorine atoms, as dictated by the usage of the FpA organic molecules when packed in the 2D lattice, play an important role to the whole structure and its properties, both electrical and optical. Last, the phenethylamine analogue system of the one presented here, did not provide emission of light although it possesses equally strong free exciton PL.

In the supplementary section, an I-V characteristic curve can be found, for a LED based on (FpA-H)$_2$PbI$_4$, which up to c.a. 5.5V displays diodic behavior (Fig. S5). The diode quality factor, for the part that the thermionic emission approximation holds, appears to be 24. It is crucial to point out that low intensity ambient light alters the I-V characteristics as the semiconductor is strongly photoconductive as well as photovoltaic. At voltages above 5.5V, the exponential behavior of I *wrt* V was not observed but rather a decrease in current flow was exhibited. While this behavior has been reproducibly verified, it may be attributed to some sort of breakdown and probably implies a relation of the quantum efficiency *wrt* to the current flow. Breakdown is visually observed with the Ga/In tip when driving at voltages larger than 11V, probably due to pinholes in the film. It is also possible that the breakdown exists as a result of some form of electrolysis that may occur on the amine salt. Research on these details as well as on the EL mechanism will be discussed in future works.

## 4. Conclusions

In conclusion, EL at low voltages is observed from lead iodide based 2D hybrid organic-inorganic semiconductor, where the organic molecule separating the 2D inorganic layers is protonated 4-fluorophenethylamine. EL is observed with naked eye at room temperature and becomes more intense at 77K. Device fabrication is afforded by one simple active layer, while no electron or hole injection layers are required. It is expected that this class of 2D low dimensional semiconductors will further advance the science of 2D hybrid organic-inorganic perovskites leading to control of complex phenomena and fabrication of advanced devices. In particular, the 2D HOIS and the quasi-2D analogues can span with excitonic peaks the spectra from the UV region and are much more tunable than the 3D HOIS, due to the combined effect of the structure, stoichiometry and dielectric confinement on the position of the excitonic peak.

**Figure 1**

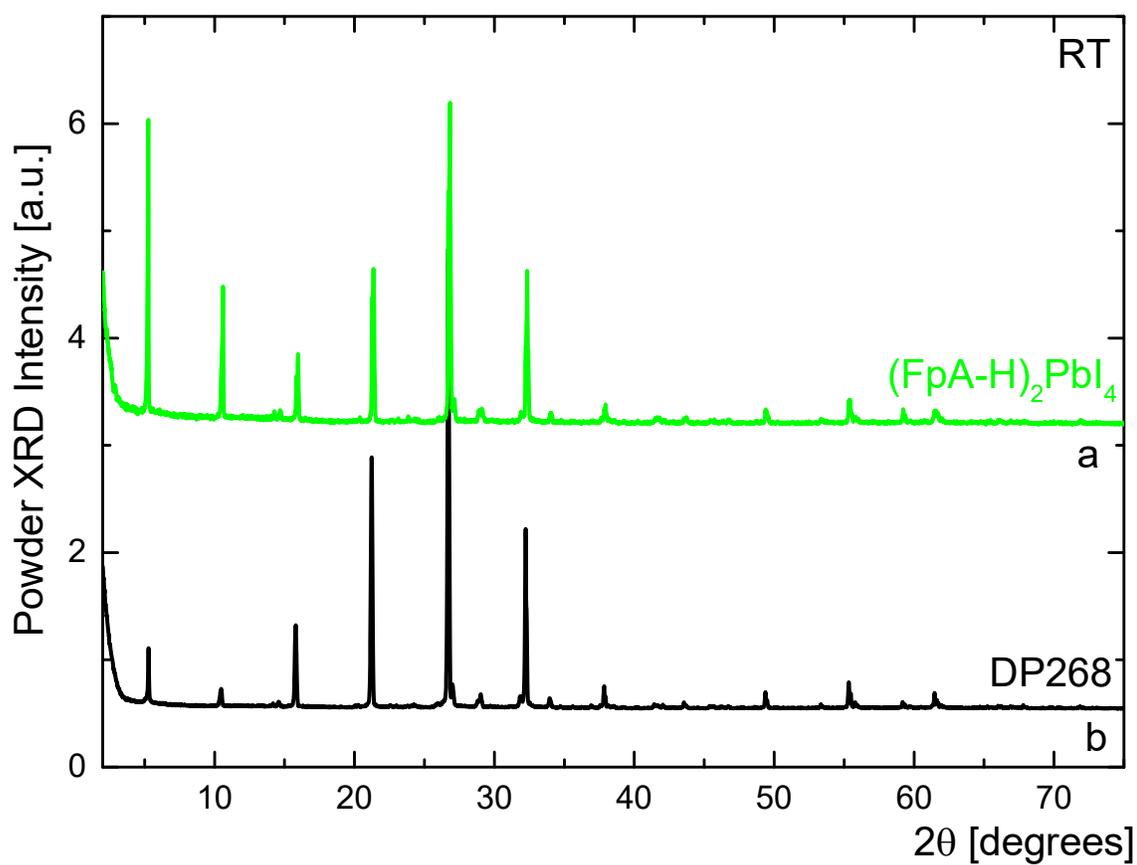

**Figure 1.** Powder XRD patterns of (a) (4-fluorophenethylamine-H)$_2$PbI$_4$ and (b) DP268 in form of thick films deposited on substrates. DP268 is the product of the (4-fluorophenethylamine-H)$_2$PbI$_4$ synthesis where the initial precursor molar ratios of amine:acid:PbI$_2$ was 8:6.3:1 instead of 2:4.3:1 as in (a).

**Figure 2**

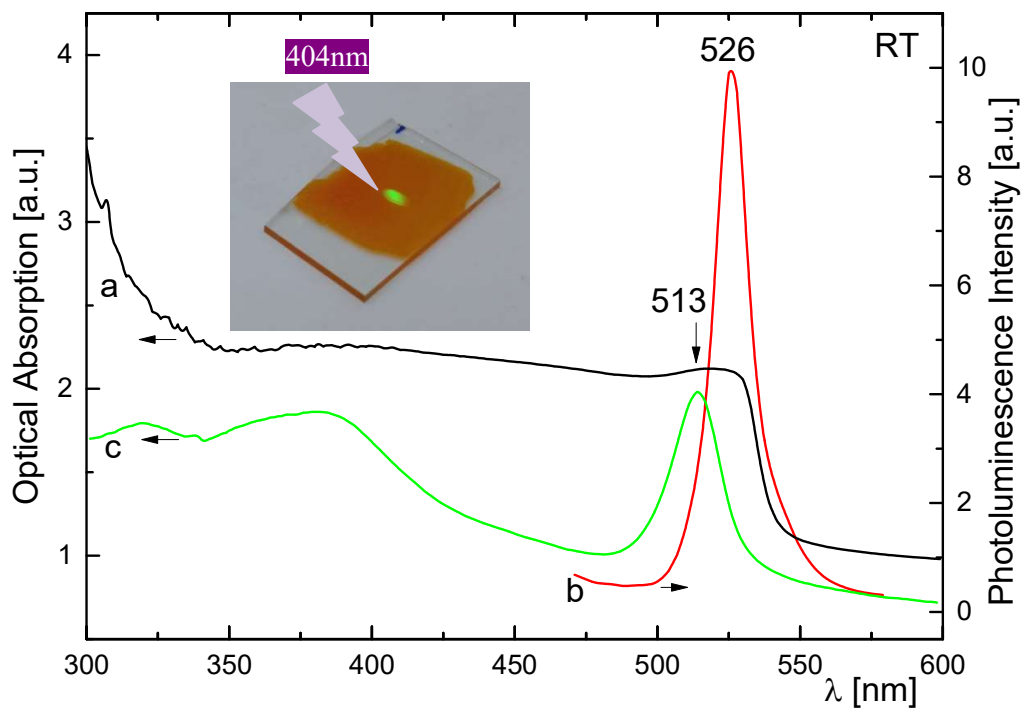

**Figure 2.** (a) Optical absorption and (b) photoluminescence spectra with ($\lambda_{exc}$=300nm), for an ITO glass substrate coated with (4-fluorophenethylamine-H)$_2$PbI$_4$ (seen in inset). (c) Optical absorption of the same material in the form of thin film as deposited on quartz substrate. Inset shows the sample depicted in (a), (b) being illuminated at the center with a 404nm point laser.

**Figure 3**

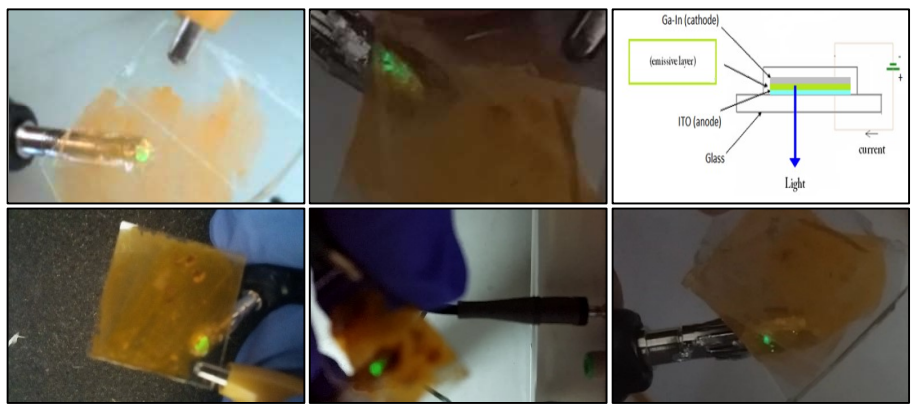

**Figure 3.** Electroluminescence images from various LEDs fabricated by coating (4-fluorophenethylamine-H)$_2$PbI$_4$ on ITO glass. Light intensity varies depending on the thickness of the film and the Ga/In contact area. Top right image shows the device schematic used.

**Figure 4**

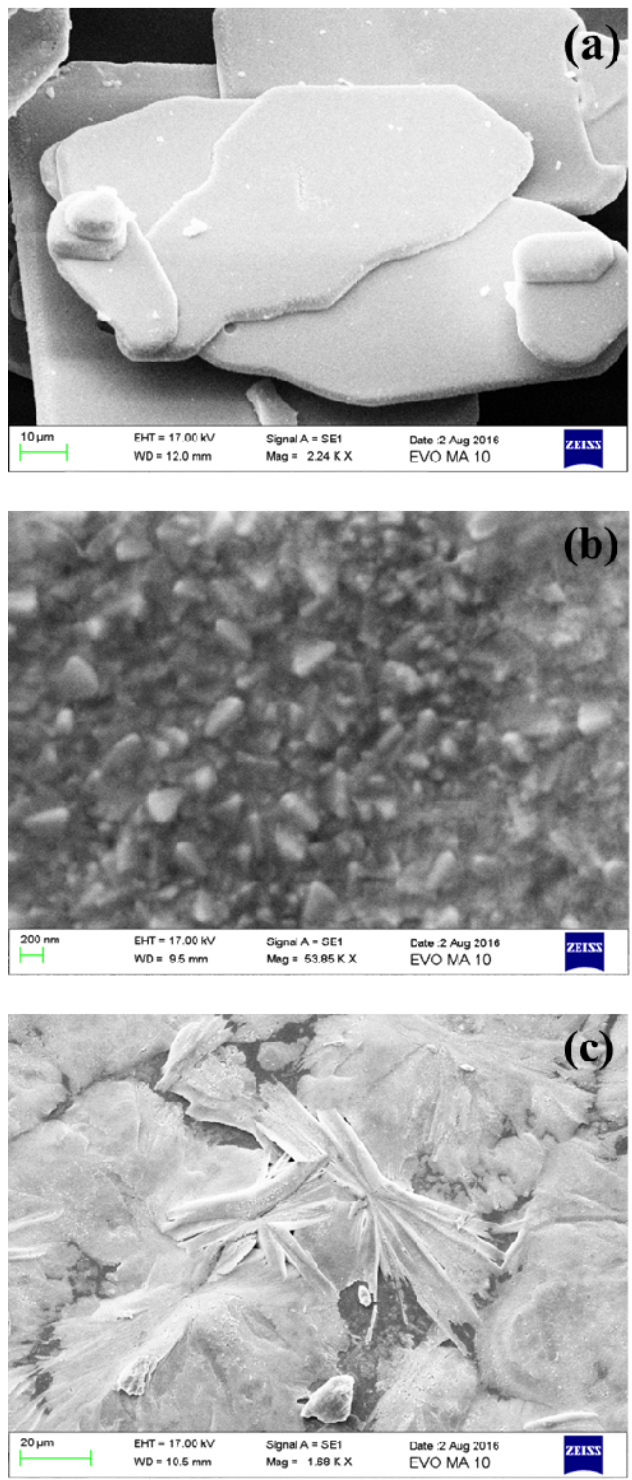

**Figure 4.** SEM images of (4-fluorophenethylamine-H)$_2$PbI$_4$: (a) in powder form synthesized from acetonitrile, (b) and (c) as deposited on heated ITO substrates from heated dimethylformamide solution. The primary observed (a) plates are transformed on ITO (b) to an array of much smaller randomly connected thin orthogonal-like plates.

**Supplementary Information S1**

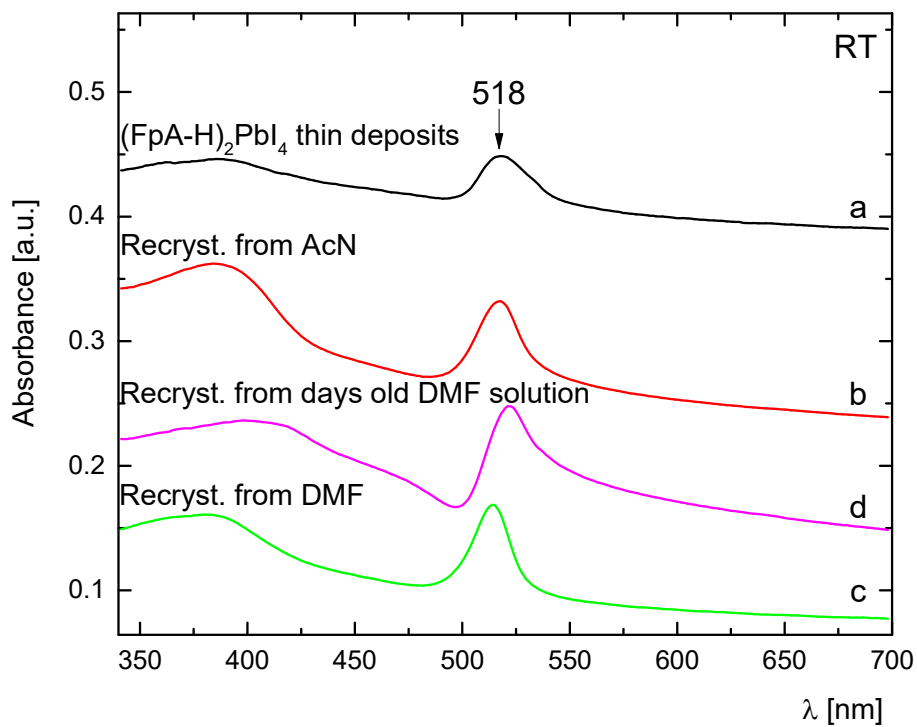

**Figure S1.** Optical absorption spectra for thin deposits of (4-fluorophenethylamine-H)$_2$PbI$_4$ of (a) dry powder synthesized in acetonitrile and stored for long period, (b) as recrystallized from acetonitrile, (c) as recrystallized from dimethylformamide and (d) as recrystallized from dimethylformamide solution which was stored for days under air before spectroscopic measurements.

**Supplementary Information S2**

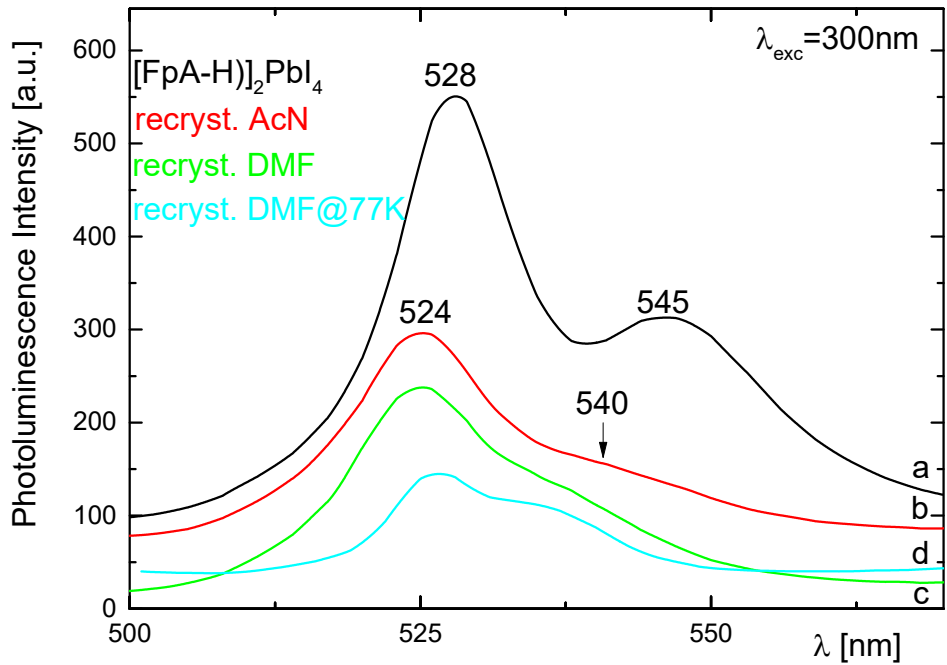

**Figure S2**. Photoluminescence spectra ($\lambda_{exc}$=300nm) for (a) thin deposits of (4-fluorophenethylamine-H)$_2$PbI$_4$ as synthesized from acetonitrile and stored for long period, (b) as recrystallized from acetonitrile, (c) as recrystallized from dimethylformamide and (d) as recrystallized from dimethylformamide and measured at 77K. Spectra (a), (b), (c) were measured at RT.

**Supplementary Information S3**

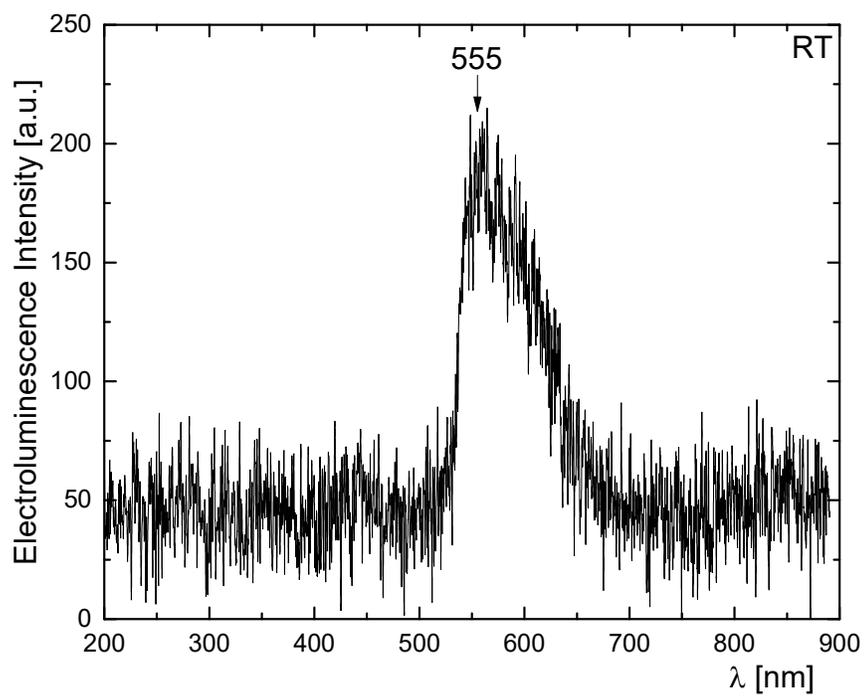

**Figure S3**. Electroluminescence spectra of a LED device based on (4-fluorophenethylamine-H)$_2$PbI$_4$ re-crystallized as thin film from heated dimethylformamide solution and deposited on a heated ITO substrate. Spectra has been acquired at RT.

**Supplementary Information S4**

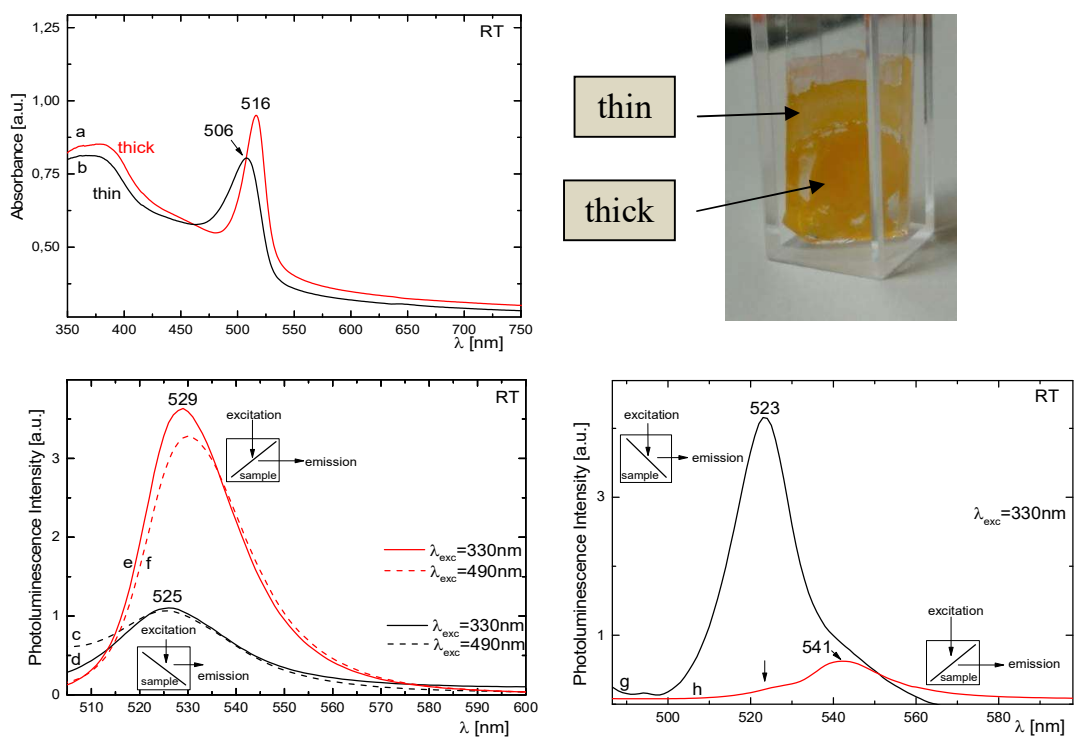

**Figure S4**. Optical absorption (OA) and photoluminescence (PL) spectra for two identical films of (4-fluorophenethylamine-H)$_2$PbI$_4$, except for their thickness. Image at top right shows such two different films on a single quartz substrate, the thin being at the outer perimeter and the thick in the inner part of the quartz substrate. PL spectra have been acquired in two configurations, a front-face (FF) and back-face (BF), as shown in the inset schematics. Spectra are: (a) OA of thick film, (b) OA of thin film, (c) PL of thin film at $\lambda_{exc}$=330nm and FF configuration, (d) PL of thin film at $\lambda_{exc}$=490nm and FF configuration, (e) PL of thin film at $\lambda_{exc}$=330nm and BF configuration, (f) PL of thin film at $\lambda_{exc}$=490nm and BF configuration, (g) PL of thick film at $\lambda_{exc}$=330nm and FF configuration and (h) PL of thick film at $\lambda_{exc}$=330nm and BF configuration. The most interesting effect is seen in (g) and (h) spectra where self-absorption leads to an apparent shift of the PL peak by 20nm.

**Supplementary Information S5**

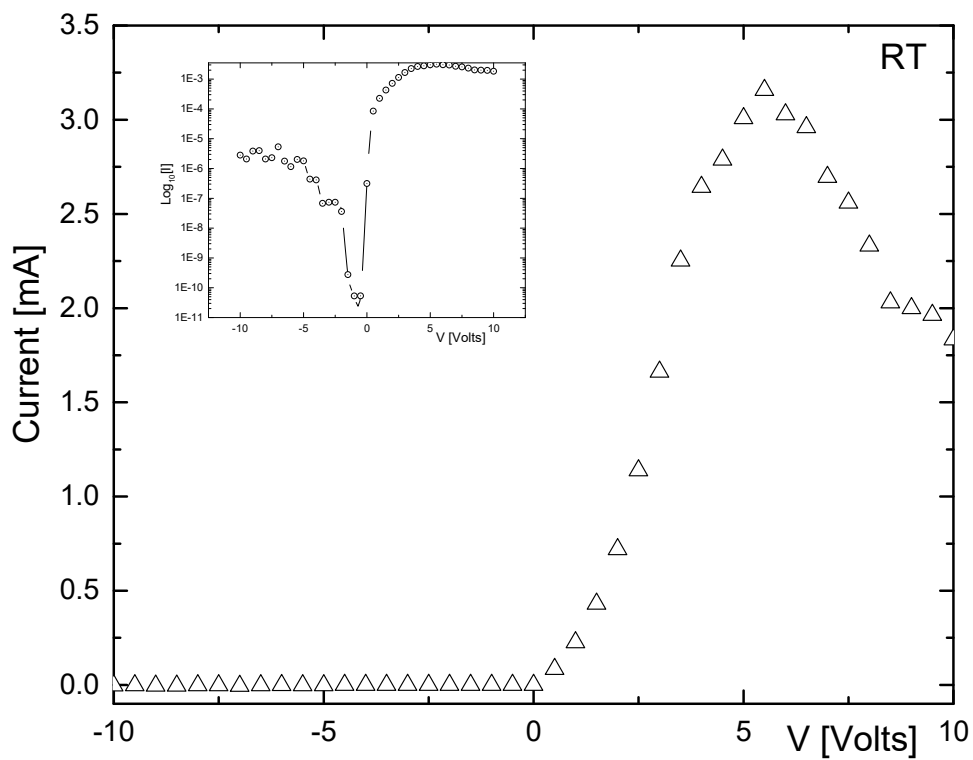

**Figure S5**. I-V characteristic curve for a LED device based on (4-fluorophenethylamine-H)$_2$PbI$_4$, which was originally synthesized in acetonitrile, re-crystallized as film from dimethylformamide and measured at RT. Positive biasing (V) occurs when the Ga/In tipped copper wire on the surface of the semiconductor is negative *wrt* to the ITO substrate.